\documentclass[12pt]{article}
\usepackage{graphicx} 
\usepackage{times} 
 \usepackage{mathptmx} 
\usepackage{setspace} 
\usepackage{graphicx}
\usepackage{xcolor}
\usepackage{hyperref}
\usepackage{booktabs}
\usepackage{svg}
\usepackage{cite}
\usepackage{amsmath}
\usepackage{bm}
\usepackage{amssymb}

\title{Self-supervised Learning for Pathological Speech Detection
}
\author{Shakeel A.~ Sheikh \\ \textit{Idiap Research Institute, Martigny, Switzerland}}
 
\begin{document}

\maketitle
\begin{abstract}
   Speech production is a complex phenomenon, wherein the brain orchestrates a sequence of processes involving thought processing, motor planning, and the execution of articulatory movements. However, this intricate execution of various processes is susceptible to influence and disruption by various neurodegenerative pathological speech disorders, such as Parkinson’s disease, resulting in dysarthria, apraxia, and other conditions. These disorders lead to pathological speech characterized by abnormal speech patterns and imprecise articulation. Diagnosing these speech disorders in clinical settings typically involves auditory perceptual tests, which are time-consuming, and the diagnosis can vary among clinicians based on their experiences, biases, and cognitive load during the diagnosis. Additionally, unlike neurotypical speakers, patients with speech pathologies/impairments are unable to access various virtual assistants such as Alexa, Siri, etc.  To address these challenges, several automatic pathological speech detection approaches have been proposed. These approaches aim to provide efficient and accurate detection of speech disorders, thereby facilitating timely intervention and support for individuals affected by these conditions. These approaches mainly vary in two aspects: the input representations utilized and the classifiers employed. 
   Due to the limited availability of data, the performance of detection remains subpar.
Self-supervised learning (SSL) embeddings, such as wav2vec2, and their multilingual versions, are being explored as a promising avenue to improve performance. These embeddings leverage self-supervised learning techniques to extract rich representations from audio data, thereby offering a potential solution to address the limitations posed by the scarcity of labeled data. Integrating self-supervised learning embeddings into pathological speech detection approaches could lead to more robust and accurate models, capable of handling diverse speech patterns and variations encountered in real-world scenarios.
 
\end{abstract}

\section{Introduction}
Speech production is a complex process which involves the coordination of various physiological functions and anatomical structures necessary to produce the intended speech sounds. This coordination of functions goes through several stages including conceptualisation, linguistic encoding, motor planning, motor execution and articulation~\cite{speechprod}. In conceptualisation, the speaker first formulates the desired speech by conceptualising ideas and organising them into coherent meaningful sentences. After conceptualisation, a linguistic encoding of these ideas is required by accessing mental lexicons and selecting grammatically appropriate word units. Once the encoding is complete, the brain plans the coordinated sequence of movements of the articulators including jaw, palate, vocal folds, lips, etc., These coordinated sequence of movements are then executed in the motor execution stage where a brain sends a sequence of signals to the muscles involved in speech production. After the transmission of brain signals to various muscles, precise movements of these muscles/articulators are required to produce various speech spounds necessary for smooth communication~\cite{damico2010handbook}. However, the execution of these various processes can be influenced by various neurodegnerative impairments. Among these neurological conditions,  Parkinson's disease (PD) or Amytrophic Lateral Sclerosis impairs the speech production mechanism leading to dysarthria, stuttering, cluttering, etc., and may have an impact on the patient's overall ability in-terms of imprecise articulation, insufficient prosody and other abnormal speech patterns~\cite{tjaden08, yunusova2008articulatory}. This speech disorder "dysarthria" can be one of the earliest indications of PD and its precise diagnosis is extremely crucial in clinical settings~\cite{darley1969differential, yunusova2008articulatory}. In addition, perceptive evaluation of pathological speech is extremely time consuming and is inclined towards the subjective belief and implicit biases of clinicians. Additionally, unlike neurotypical speakers, patients with speech pathologies/impairments are unable to access various virtual assistants such as Alexa, Siri, etc~\cite{s2022machine}. To assist in clinical diagnosis of pathological speech evaluation, several automatic methods have been proposed to provide time efficient, cost effective and objective assessment of pathological speech. These automatic methods broadly fall in two categories including traditional classical machine learning employing hand-crafted features and deep learning employing mostly time-frequency input features.  
In traditional machine learning  approaches, handcrafted features~\cite{gillespie2017cross, kodrasi2020automatic, kodrasi2020spectro, s2022machine} inspired by clinical knowledge are fed to classical algorithms such as support vector machines (SVMs) or logistic regression to discriminate between healthy and atypical pathological speech. 
Using such specific features with simple machine learning models yields interpretable results, which is critical in a clinical context. However, their predictive power are limited and doesn't outperform the expertise of top human professionals~\cite{parvanehthesis, sheikh2023deep}.
\par 
Conversely, deep leaning models rely more on data. These models directly analyze transformed data such as spectrograms or Mel-frequency cepstral coefficients without prior feature extraction but require more complex architectural components (e.g., convolutional neural networks~\cite{janbakhshi_stft, s2023stuttering, sheikh2021stutternet, s2023advancing, VASQUEZCORREA202056, sheikh2022robust, narendra2021detection, vaiciukynas2018parkinson}, long short-term memory networks~\cite{mayle2019diagnosinglstm}, autoencoders~\cite{janbakhshi2021supervised, janbakhshi2022adversarial, vasquez2017convolutional}, etc.) and more data to be trained. As a result, they often achieve significantly higher performance. To this end, several deep leaning models have been explored for pathological speech domain. Deep leaning exploits data driven approaches to learn abstract pathological cues and improves the state-of-the-art performance in PD classification tasks remarkably~\cite{vasquez2017convolutional,mallela2020raw,janbakhshi_stft, janbakhshi_ua}. Recently, adversarial pathological speech detection models have been proposed to learn robust pathological cues that are speaker invariant but at the same time are pathology discriminant~\cite{janbakhshi2021supervised}. Even though this method shows good performance, however, adversarial training is unstable and is usually very sensitive to training parameters, thus making its training very challenging. To address this, P. Janbakshi \emph{et al.}~\cite{janbakhshi2022adversarial} proposed adversarial-free training where  they employ feature separation framework relying on mutual information minimization to learn speaker invariant features. Even though this shows promising results, the key challenge in guiding deep learning models to capture and extract abstract pathological cues is still limited by the availability of large pathological datasets. Moreover, with these low resource datasets, it is hard to capture various speaker and linguistic attributes interns of speaking style, gender, phonetic content, prosody and other pathology related para-linguistic cues. 

 As a result, powerful data-driven approaches, such as self-supervised learning models like wav2vec2 (w2v2), have been recently exploited~\cite{baevski2020wav2vec}. 
 These models leverage a vast collection of non labelled available audio data, learning embeddings which enable unprecedented performance for several downstream tasks~\cite{superb}. Motivated by this, several attempts have been made in adopting w2v2 models to pathological speech detection~\cite{s2022end, sheikh2022introducing, s2023stuttering, w2v2_pd, javanmardi2023wav2vec, getman2022wav2vec2}. However most of the studies employ English pre-trained version of w2v2, this inhibiting the learning of various pathological cues in diverse languages. In this study, we aim to evaluate and analyze the performance of w2v2 and its multilingual variant and provide in depth analysis using the embedding layers of transformer block for pathological speech detection. 

\section{Contextual Embeddings}
The w2v2 self-supervised learning model is a speech recognition model and is comprised of three blocks including  feature encoder, quantization and transformer contextual block. The feature encoder  $\mathbf{f}: \mathbf{X}\rightarrow \mathbf{Z}$ converts raw audio waveform into local feature representations  $\mathbf{Z}= \mathbf{f}(\mathbf{X})$ (with $\mathbf{Z}$ = [$\mathbf{z}_1$, $\mathbf{z}_1$,..., $\mathbf{z}_T$] are temporal features) and contains a stack of 1D convolutional, batch norm and GELU activation layers. These features are then passed to a contextual transformer block $\mathbf{h}: \mathbf{Z}\rightarrow \mathbf{C}$ to learn meaningful contextual embeddings $\mathbf{C}= \mathbf{h}(\mathbf{Z})$. The contextual block $\mathbf{C}$ is comprised of 24 attention layers with each layer consists of 24 self attention heads. This self-attention mechanism allows the w2v2 model to learn and capture long contextual dependencies from the input audio sequence. The encoded feature representations $\mathbf{Z}$ are also passed to quantization module $\mathbf{q}: \mathbf{Z}\rightarrow \mathbf{Q}$ that maps continuous-valued local representations into a set of discrete quantized codes by applying vector quantization methods like k-means clustering. The quantization module $\mathbf{Q}$ consists of two (320 possible entries) code books. For each local representation $\mathbf{z}_i$ $\in$ $\mathbf{Z}$, a code of 320-dimensional is chosen from each code book and is then concatenated afterwards. This is followed by a linear transformation to obtain $\mathbf{q}_i$ $\in$ $\mathbf{Q}$ vectors. The code is chosen using 

\begin{equation}
    p_{g,v} = \frac{\exp(l_{g,v} + \eta_v)/\tau}{\sum_{k=1}^{V}\exp(l_{g,v} + \eta_v)/\tau}
    \label{gumbel}
\end{equation}
where  $v$ is v-th codebook entry, $l$ is logit, $g$ is codebook group, $\eta=-\log(-log(u))$ with $u$ are uniform samples from $\mathcal{U}(0,1)$, and $\tau$ represents temperature which controls the randomness.  The model is trained in a self-supervised fashion to learn very rich contextual embeddings by optimizing the loss function:

\begin{equation}
    \mathcal{L} = \mathcal{L}_c + \alpha\mathcal{L}_d
\end{equation}
where $\mathcal{L}_c$ is contrastive objective loss given by 
\begin{equation}
    \mathcal{L}_{c} = -\log\frac{\exp(sim(c_t, q_t)/\tau)}{\sum_{\tilde{q}\in Q}\exp(sim(c_t, \tilde{q})/\tau)}
    \label{eq:contrastiveloss}
\end{equation}
where $sim(c_t, q_t)$ is the cosine similarity between the $q_t$ and $c_t$. The $\mathcal{L}_d$ is diversity loss given by 
\begin{equation}
    \mathcal{L}_d = \frac{1}{GV}\sum_{g=1}^G-H(\hat{p}_g) = \frac{1}{GV}\sum_{g=1}^G\sum_{v=1}^V\hat{p}_{g,v}log\hat{p}_{g,v}
\end{equation}
where $g$ is the $g^{th}$ codebook group and $v$ is the $v^{th}$ codebook entry. The authors have released several variants of w2v2 models. However, we use large and multilingual   XLRS-53 variant in our case study. 
\par 
The SUPERB bechmark~\cite{superb} has shown remarkable progress of w2v2 self-supervised learning embeddings in various downstream speech applications~\cite{pepino2021emotion, vsvec2022evaluation, janbakhshi_ua, s2023stuttering, s2022end, w2v2_pd}. Motivated by this, we extract and exploit embeddings from XLRS-53 multilingual model~\cite{conneau2020unsupervised}. The XLRS-53 is trained on a massive amount of 56K hours of data using multiple datasets including multilingual LibriSpeech, CommonVoice and BABEL.  It has been experimentally proven that the middle layers show state of the art performance in various pathological speech disorders~\cite{sheikh2022introducing, s2023stuttering}. In this study, we analyse, evaluate and compare the performance of each layer separately by applying a linear layer as a classification head to reveal the prediction class.

\section{Methodology}
\subsection{Dataset}
All the experimental analysis are performed on Colombian Spanish PC-GITA dataset~\cite{orozco2014new}. The PC-GITA dataset comprises of 100 speakers with 50 healthy speakers and 50 pathological speakers diagnosed with Parkinson's disease. For this case study, we use phonetically balanced recording of 10 sentences and readspeech. The age range of male PD speakers is 33 to 77 years old with a mean of 62.2 $\pm$ 11.2, and the age range of female PD speakers is 44 to 75 years old with a mean of 60.1 $\pm$ 7.8. The dataset is well balanced with neurotypical speakers having a male age ranges of 31 to 68 with a mean of 61.2 $\pm$ 11.3 and female average range 60.7 $\pm$ 7.7. The audio recordings were collected in a sound proof noise controlled environment with a sampling rate of 44.1 kHz. 
\subsection{Implementation}
For our considered experimental studies, we use PyTorch and Torchaudio as implementation tools. All the considered approaches are trained using Adam optimizer and the model weights are initialized randomly using Xavier initialisation via 
\begin{equation}
    W_{ij} = \mathbf{U} \Biggr[ -\frac{\sqrt{6}}{\sqrt{n_{in} + n_{out}}}, \frac{\sqrt{6}}{\sqrt{n_{in} + n_{out}}}\Biggr]
\end{equation}
where $\mathbf{U}$ is the uniform distribution and $n_{in}, n_{out}$ are the number of input and out neurons respectively. The initial learning rate is set to $10^{-2}$ and is reduced after every 15 iterations using the PyTorch MultiStepLR scheduler with $\gamma$ = 0.9. We terminate the training using the early stopping criteria and is terminated if the validation loss doesn't reduce for 10 successive iterations. For inference at test time, we use the last best saved model for evaluation purposes. The results reported in this chapter are evaluated on the 10-fold validation strategy i.e., for each fold, samples from 80\% speakers of speakers are used for training, samples from 10\% of remaining speakers are used for validation and the samples from the remaining 10\% speakers are used for testing purposes. 
Since the dataset is class balanced, we use accuracy metric to be consistent with the literature. We report speaker level performance on unseen test speakers via soft voting on prediction scores given by:  
\begin{equation}
   \text{Soft Voting} = \frac{1}{\mathbf{N}(S_k)}\sum_{i = 1, k \in \mathbf{S}}^{\mathbf{N}(S_k)} p^{k}_i
\end{equation}
where $p_i^k$ is the $\mathbf{R}^2$-dimension prediction probability score of a sample $i$ of speaker $S_k$, and $\mathbf{N}(S_k)$ is the total number of samples from speaker $S_k$.
 \subsection{Embedding Extraction}
 For embedding extraction of each sample of PC-GITA dataset, we extract $\mathbf{R}^{ L \times 768 \times T}$, where $L$ represents the layer embedding and $T$ is temporal dimension of each input sequence. After extracting features, we apply statistical (mean and standard deviation) pooling across temporal domain on each sample $\mathbf{R}^{ L \times 768 \times T}$ to get a fixed $\mathbf{R}^{ L \times 2 \times T}$-dimension embedding vector. This $\mathbf{R}^{ L \times 2 \times T}$-dimension vector is then fed to a downstream classifier for pathological speech detection.  

         

\begin{table}
    \centering
    \begin{tabular}{c|c|c|c}
    \toprule
    Layer&\multicolumn{3}{c}{Accuracy} \\
    \midrule
       & w2v2 large & w2v2 XLRS-53 & XLRS-53 (Spanish) \\
       \midrule
            1&	77.0&	77.0 & 78.0 \\
            2&	77.0&	76.0 & 74.0   \\
            3&	78.0&	77.0  & 77.0  \\
            4&	78.0&	85.0  & 80.0  \\
            5&	80.0&	82.0  & 80.0  \\
            6&	84.0&	82.0  & 82.0 \\
            7&	81.0 &	83.0  & 81.0 \\
            8&	78.0&	84.0  & 79.0\\
            9&	80.0&	80.0  &82.0 \\
            10&	78.0&	82.0   &75.0\\
            11&	78.0&	81.0   &73.0\\
            12&	79.0&	76.0  &80.0\\
            13&	85.0&	78.0   &81.0\\
            14&	76.0&	80.0  &86.0\\
            15&	74.0&	83.0  &80.0\\
            16&	72.0&	80.0  &78.0\\
            17&	72.0&	75.0  &84.0\\
            18&	75.0&	84.0  &84.0\\
            19&	71.0&	84.0  &85.0\\
            20&	74.0&	82.0  &80.0\\
            21&	67.0&	84.0  &79.0\\
            22&	69.0&	69.0  &78.0\\
            23&	67.0&	58.0  &74.0\\
            24&	71.0&	60.0   &71.0\\
            
         \bottomrule
    \end{tabular}
    \caption{Pathological classification accuracy on PC-GITA dataset. The results in the last column are based on fine-tuning XLRS-53 w2v2 model first with Spanish dataset and then the embeddings are extracted from the Spanish trained XLRS-53 before feeding it to downstream pathological classification head.}
    \label{tab:table}
 \end{table}

\begin{figure*}
\hspace{-4cm} 
    \includegraphics[scale=0.3]{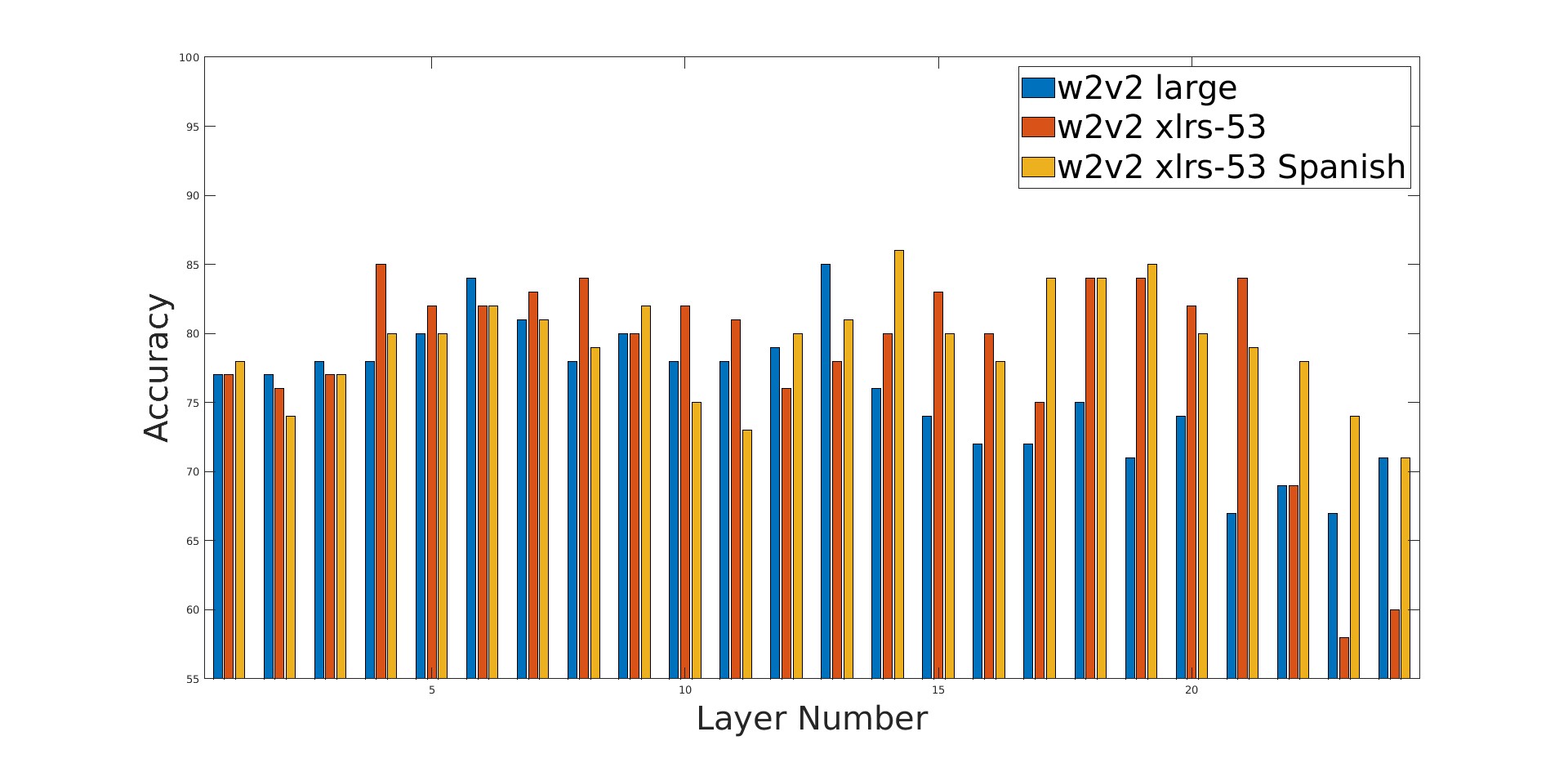}
    \caption{Pathological classification accuracy using various layers of w2v2 large and XLRS-53 self-supervised learning model. The blue one is based on w2v2 large variant trained on English audio data only. The red is based on w2v2 XLRS-53 multilingual model trained on 56K hours of audio data. The yellow is fine-tuned version of XLRS-53 fine-tuned on Common Voice 6.1 Spanish dataset.}
    \label{fig:layeracc}
\end{figure*}

\section{Results and Discussion}
Different layers of w2v2 model capture different acoustic information in terms of phonetic content, speaking style, emotions, speaker characteristics, etc. Motivated by this, we experiment and analyse the impact of all the embedding layers of w2v2 model.  Table~\ref{tab:table} presents the results using different w2v2 embedding layers using  w2v2 large model. From the results, we analyse that the initial three contextual layers of w2v2 large variant tend to show slightly better performance than the XLRS-53 variant on pathological speech detection with the peak performance at layer 13. Even though the large variant of w2v2 shows good peak performance, however the variability across different layers is not very robust, as depicted in Figure~\ref{fig:layeracc}. Overall, the middle layers show good pathological detection performance in comparison to last layers, where the performance drops very significantly. This is possible due to the reason that the large variant of w2v2 is fine towards automatic speech recognition task and it is likely possible that the para-linguistic cues necessary for pathological detection are lost. 
\par 
Table \ref{tab:table} also shows the results of various embedding layer when employing multilingual (XLRS-53) variant of w2v2 model. It can be seen from the results that the XLRS-53 embeddings exhibit consistently better performance and lesser variability in pathological speech detection across various contextual layers, thus making it robust across layer embeddings.  This is also clear from the Figure~\ref{fig:layeracc}. The reason is possibly that the XLRS-53 variant is trained on 56K hours of multilingual audio data such as multilingual LibriSpeech, CommonVoice and BABEL. This helps the XLRS-53 model to learn and extract robust features in terms of speaker characteristics, phonetic content, para-linguistic information, prosody, an so on. However, we also not that the performance within the last few layers is degrading quite significantly as compared to large variant of w2v2 model. To overcome this limitation, we first use Spanish Common Voice 6.1 dataset for fine-tuning and updating the weights of XLRS-53 w2v2 model to capture various characteristics of Spanish language. The results from Table~\ref{tab:table} and Figure~\ref{fig:layeracc} show that the fine-tuning of XLRS-53 model on Spanish dataset does indeed boost the performance of last layers and also shows consistently lesser variability across contextual layers.  

\section{Conclusion}
The pathological datasets are extremely scarce and low. To address this limitation, we exploit self-supervised learning framework where a model is first pre-trained on massive amounts of data capturing various speaking styles, linguistic content, speaker attributes, para-linguistic information and so on. To this end, we exploit and provide  the detailed analysis of self supervised learning w2v2 model and its variants in pathological speech detection. Results show that the XLRS-53 is good at capturing various pathological cues and gives state of the art performance. Moreover finetuning of XLRS-53 on Spanish dataset further boosts the performance of pathological speech detection in the last contextual layers. While self-supervised learning models demonstrate promising performance, their ability to capture linguistic and phonetic information across diverse temporal dimensions may potentially hinder the recognition of pathological cues. In future work, we will explore models that learn phonetic invariant representations with an aim to improve the performance of pathological speech detection. In addition, large language models for pathological speech correction would be an interesting idea to explore in further studies. 
\bibliographystyle{plain}
\bibliography{ref}
\end{document}